\input{aipcheck}

\documentclass[
    ,final            
  ]
  {aipproc}

\layoutstyle{8x11double}

\usepackage{axodraw,picture,epsfig}
\newcommand{\amuLH}{{a_\mu^{\rm LHV}}}
\newcommand{\bi}{{\begin{itemize}}}
\newcommand{\ei}{{\end{itemize}}}
\newcommand{\Tr}{{{\rm Tr}}}

\begin{document}
\title{The leading hadronic vacuum polarisation on the lattice}


\author{Michele Della Morte}{
  address={Institut f\"ur Kernphysik and Helmholtz Institut Mainz,
Johannes Gutenberg-Universit\"at, 55099 Mainz, Germany}
}

\author{Benjamin J\"ager}{
  address={Institut f\"ur Kernphysik, University of Mainz, Becher Weg 45, 55099 Mainz, Germany}
}

\author{Andreas J\"uttner}{
  address={CERN, Physics Department, TH Unit, CH-1211 Geneva 23, Switzerland} }
\author{Hartmut Wittig}{
  address={Institut f\"ur Kernphysik and Helmholtz Institut Mainz,
Johannes Gutenberg-Universit\"at, 55099 Mainz, Germany}
}

\begin{abstract}
After discussing the relevance of a first principles theory-prediction of the hadronic vacuum polarisation for Standard Model tests, the theoretical challenges for its computation in lattice QCD are reviewed. New ideas that will potentially improve the quality of lattice simulations will be introduced and the status of ongoing simulations will be presented briefly.
\end{abstract}

\maketitle


\section{Introduction}
\begin{figure}[b]
 \begin{picture}(120,60)(-45,-10)
 \SetScale{.8}
  \ArrowLine(-34,-12)(0,30)
  \ArrowLine(0,30)(+34,-12)
  \Photon(0,30)(0,50){2}{3}
  \SetColor{Black}
  \Text(-20,14)[c]{$\mu$}
  \Text(+6,40)[c]{$\gamma$}
  \Oval(0,0)(6,12)(0)
  \SetColor{Black}
  \Text(+17,-10)[c]{$\gamma$}
  \Text(-17,-10)[c]{$\gamma$}
  \Text(-0,12)[c]{$q$}
  \Text(-0,-12)[c]{$\bar q$}
  \Photon(-24,0)(-12,0){2}{2}
  \Photon(12,0)(24,0){2}{2}
 \end{picture}
 \caption{The leading hadronic contribution $\amuLH$.}\label{fig:LHV}
\end{figure}
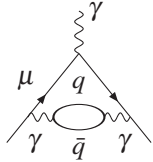

The quest to measure the anomalous magnetic moment of the muon,
$a_\mu$,  experimentally with ever increasing precision 
has started at CERN~\cite{Charpak:1961mz} decades ago and was later continued
at BNL~\cite{Bennett:2004pv}, yielding a precision of 
0.5ppm~\cite{Bennett:2006fi}. In the Standard Model (SM) $a_\mu$ receives
contributions most notably from QED but also from the Weak sector and from QCD. 
While the former two 
contributions can be computed in (high order) perturbation theory, the QCD 
contributions are non-perturbative and as such not computable analytically in a
model independent way. 
As nicely summarised in \cite{Jegerlehner:2009ry},
using unitarity and analyticity, the
leading QCD-contribution (cf. figure~\ref{fig:LHV})
is currently determined from the experimental measurement of
$e^+ e^-$-annihilation or hadronic $\tau$-decays. With these experiment-based
approaches one is able to predict $\amuLH$ with a precision below 1\%.
While the former process seems to provide solid
SM-predictions, a pure theory-prediction still
seems worth-while. In particular since $a_\mu$ 
is sensitive to contributions from physics beyond the SM. In fact, 
the theory prediction currently differs from the experimental measurement 
by 3.2$\sigma$ \cite{Jegerlehner:2009ry}. While not yet providing evidence
for new physics, this tension is naturally causing excitement.
 
Here we report on progress in predicting $\amuLH$
in lattice QCD. It will become clear that this type of 
calculations is still in its infancy 
\cite{Gockeler:2003cw,Aubin:2006xv,Renner:2010zj,Brandt:2010ed}. 
To this end the precise experimental
prediction of $a^{\rm LHV}_\mu$ will in the near future remain a reference for 
assessing lattice computations.\\[-8mm]
\section{$\amuLH$ on the lattice}\mbox{}\\[-4mm]
The leading hadronic contribution is defined as
\begin{equation}\label{eq:master}
\amuLH =\left(\frac{\alpha}{\pi}\right)^2
        \int\limits_0^\infty dQ^2 K(Q^2){\left(\Pi(Q^2)-\Pi(0)\right)},
\end{equation}
where the function $K(Q^2)$ parametrises the QED contribution to the diagram in 
figure \ref{fig:LHV} \cite{Lautrup:1971jf}. Note that the integration is over
space-like momenta $Q^2$. The function 
$\Pi(Q^2)$ is the vacuum polarisation amplitude,
\begin{eqnarray}\label{eq:VPtensor}
  \Pi_{\mu\nu}(q)&=&\int d^4 x e^{\,iq(x-y)}
        \langle\, j^{\rm EM}_\mu(y)j^{\rm EM}_\nu(x)\rangle\nonumber\\[-2mm]
	\\[-2mm]
	&=&(q_\mu q_\nu - q^2 g_{\mu\nu}){\Pi(Q^2)}\,,\nonumber
\end{eqnarray}
where $Q^2\equiv -q^2$.
Simulations of lattice QCD always have to keep track of a number of systematic 
effects, most notably those stemming from the finite lattice-spacing, 
the finite volume and unphysically heavy quark masses. These effects have 
been studied deeply and a large
body of tools by now allows one to estimate or to systematically
control them at the level of precision required here. In this talk 
we present new ideas
tailored to control systematic uncertainties arising  from 
the presence of quark-disconnected diagrams, from the limited 
momentum resolution in finite volume field theory and from the contribution of 
vector resonances.

\noindent{\bf Quark disconnected diagrams and vector resonances}\\
The electro-magnetic current $j_\mu^{\rm EM}(x)$ in eq.~(\ref{eq:VPtensor}) 
consists of 
a linear combination of flavour-diagonal quark-bilinear currents 
$\bar q(x)\gamma_\mu q(x)$, where $q=u,d,s$.
The Wick contractions of eq.~(\ref{eq:VPtensor}) therefore yield quark connected 
contributions
\begin{equation}\label{eq:conn1}
\langle\Tr\{S_q(x,y)\gamma_\mu S_q(y,x)\gamma_\nu\}\rangle \,,
\end{equation}
and quark disconnected contributions
\begin{equation}\label{eq:disc1}
\langle\Tr\{S_q(y,y)\gamma_\mu\}\Tr\{S_q(x,x)\gamma_\nu\}\rangle\,,
\end{equation} 
where $S_q(x,y)$ is the quark propagator for the quark flavour $q$ and
the trace is over spin- and colour-indices. 
In lattice simulations the latter contribution is often neglected because
of the huge overhead its computation causes \cite{Neff:2001zr}.

A new method for predicting 
correlation functions consisting of quark-disconnected contributions
was developed in \cite{DellaMorte:2010aq}.
By  introducing valence quarks which are degenerate
with the dynamical flavours, each Wick contraction can be rewritten in terms
of a single fermionic correlation function defined in an unphysical
theory. The
physical result is recovered by summing over the correlation functions
in the unphysical, partially quenched, theory
\cite{Bernard:1992mk,Bernard:1993sv,Sharpe:2000bc}. 
In particular, the above expectation values eq.~(\ref{eq:conn1}) and 
(\ref{eq:disc1}) remain unchanged if one of the quarks is replaced by a 
mass-degenerate partially quenched valence quark $q^\prime$. This allows to  
express them in terms of individual two-point functions, the Wick-contractions
of which yield either a connected or a disconnected two-point function:
\begin{eqnarray}\label{eq:cd}
C^{\rm conn}(y,x)\equiv\langle \bar q(y)\gamma_\mu q^\prime(y) \bar q^\prime(x)\gamma_\nu q(x)\rangle\nonumber\,,\\[-2mm]
\\[-2mm]
C^{\rm disc}(y,x)\equiv\langle \bar q^\prime(y)\gamma_\mu q^\prime(y) \bar q(x)\gamma_\nu q(x)\rangle\nonumber\,.
\end{eqnarray}
Within partially quenched chiral perturbation theory 
\cite{Bernard:1992mk,Bernard:1993sv,Sharpe:2000bc,Gasser:1983yg,Gasser:1984gg},
expressions for the connected and the disconnected contributions to hadronic
correlation functions can be computed. The case
of the $N_f=2$-theory without and with a partially quenched strange quark as well 
as the $N_f=2+1$-theory were studied in \cite{DellaMorte:2010aq}. In the $N_f=2$-theory
for example, the calculation at next-to-leading order in the effective theory
predicts that the disconnected contribution reduces the connected contribution by
only 10\%. 

The analytical prediction of quark-disconnected diagrams is work in progress. In
particular, vector resonances which are not dynamical degrees of freedom in 
chiral perturbation theory turn out to be  dominating $\Pi(Q^2)$
\cite{Aubin:2006xv,Renner:2010zj}. 
It will be interesting to study the impact of vector resonances on the above
predictions by including these degrees of freedom into the chiral 
Lagrangian~\cite{Ecker:1988te}.
However, 
while the effective theory for pions and kaons stands on solid grounds, this 
similar ansatz for vector mesons is a model.

\begin{figure}
 \epsfig{scale=.25,angle=-90,file=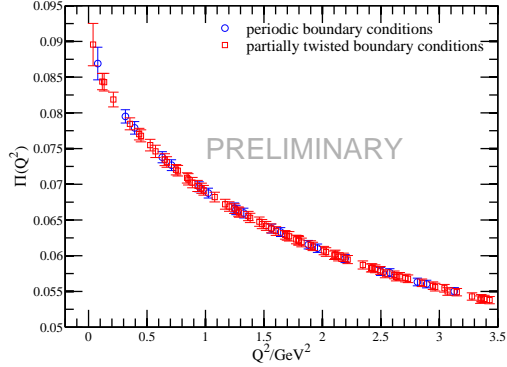}
 \caption{Results for $\Pi(Q^2)$ with (red) and without (blue) twisted
	 boundary conditions.}\label{fig:mainzres}
\end{figure}

\noindent{\bf Momentum resolution}\\
Today, typical lattices extend over $L\approx 3$fm in the spatial directions 
(typically twice that large in the temporal direction, $T=2L$). 
\begin{figure}
\epsfig{scale=.25,angle=-90,file=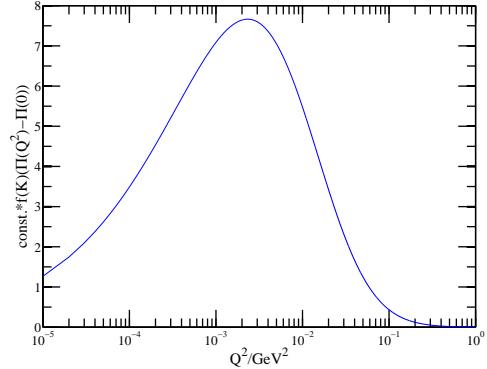}
\caption{Shape of the integral kernel in eq.~(\ref{eq:master}) assuming 
vector dominance for $\Pi(Q^2)$.}\label{fig:kernel}
\end{figure}
Besides vanishing momentum, the lowest hadron momentum therefore
corresponds to the lowest non-vanishing Fourier-mode, $2\pi/T\approx 200$MeV. 
The plot in figure~\ref{fig:kernel} shows the integral
kernel of eq.~(\ref{eq:master}), which is peaked at around the muon mass,
$m_\mu\approx 106$MeV. 
The lattice data therefore needs to be 
extrapolated into this region. 
In order to compute $\Pi(Q^2)$ closer to the peak,
the Mainz group \cite{Brandt:2010ed} 
is applying twisted boundary conditions
to the valence quark fields, $q(x+L\hat i)=e^{i\theta_i}q(x)$, 
which allows one to tune the offset of the Fourier Modes accessible
in the lattice computation 
\cite{Bedaque:2004kc,deDivitiis:2004kq,Sachrajda:2004mi,
Bedaque:2004ax,Tiburzi:2005hg,Flynn:2005in,Guadagnoli:2005be}. 
Naively, the twists applied to the quark-fields in the flavour-diagonal
currents in eq.~(\ref{eq:VPtensor}) will
cancel (see e.g. \cite{Sachrajda:2004mi}). However, the 
connected correlation function in eq.~(\ref{eq:cd}), $C^{\rm conn.}(y,x)$, is
composed of flavour-off-diagonal currents. Hence, different twist-angles can
be applied for the quark fields $q$ and $q^\prime$, respectively \cite{DellaMorte:2010aq}. This argument
allows to apply partial twisting at least to the connected contribution
to $\amuLH$. 
The disconnected contribution can either be predicted using chiral 
perturbation theory (cf. above) or it can be computed for the 
usual Fourier momenta and then be interpolated using the ansatz provided by 
chiral perturbation theory.
Figure \ref{fig:mainzres} shows results for $\Pi(Q^2)$ by the Mainz group.
Without twisting only the blue data points are accessible, while with 
twisting the red
data points can be added. Besides providing data points closer to the region
where the integral eq.~(\ref{eq:master}) receives major contributions, the
additional data points for larger values of the momentum will help in stabilising fits:
In order to compute the result for the vacuum polarisation tensor eq.~(\ref{eq:VPtensor})
one first fits an ansatz for its momentum-dependence to the data and then 
integrates it. The Mainz group uses
models for the vector resonance, polynomials and Pad\'e approximations
as ans\"atze (cf. also ETM's study of parametrisations of vector resonances 
in \cite{Renner:2010zj}). An estimate of the systematic
uncertainties is obtained from the spread of the respective fit-results. 
In our analysis the additional data points which we obtained by using 
partially twisted boundary conditions helped to reduce this spread significantly.\\[-8mm]
\section{Status}
\begin{figure}
\epsfig{scale=.25,angle=-90,file=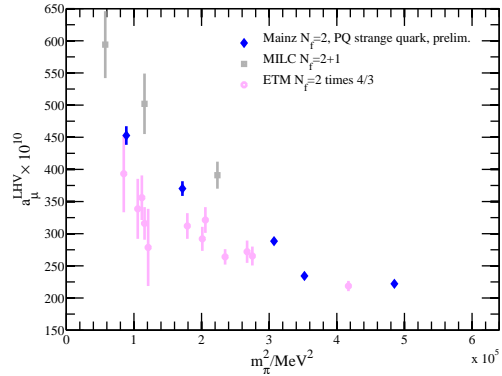}
\caption{Comparison of results for $\amuLH$ as a function of the pion mass squared: 
Sources: our results (blue diamonds) are preliminary, 
MILC \cite{Aubin:2006xv} (where we have chosen as the error bar the maximum
spread between results from different $q^2$-parametrisations) and ETMC \cite{Renner:2010zj}.
Note that the plotted results are for different values of the lattice cut-off and also
of the lattice volume.
}\label{fig:results}
\end{figure}

\mbox{}\\[-6mm]
Figure~\ref{fig:results} shows results for the vacuum polarisation by 
ETM \cite{Renner:2010zj}, MILC \cite{Aubin:2006xv} 
and by the Mainz group. Note that Mainz is the only group using
twisted boundary conditions while ETM is the only collaboration trying
to compute the disconnected contribution directly. For comparison we 
quote the value obtained from the analysis of $e^+ e^-$-annihilation, 
$\amuLH=690(5)\times 10^{-10}$ \cite{Jegerlehner:2009ry}. Clearly the lattice 
data tend towards
this value as the quark mass is reduced, but the extrapolation to the physical
point is model-dependent and will only be dispensable once the simulation is
carried out for physical quark masses. 

In this talk we present the status of lattice computations of 
$\amuLH$. While significant progress has been made recently systematic 
uncertainties are still not under sufficient control. Clearly, simulations
at the physical point would be very desirable in order to be independent of
model-based extrapolations of lattice data.\\[-8mm]
\begin{theacknowledgments}
We thank our colleagues within the CLS project for sharing gauge ensembles.
Calculations of correlation functions were performed on the dedicated QCD platform ``Wilson'' at
the Institute for Nuclear Physics, University of Mainz. 
\end{theacknowledgments}



\bibliographystyle{aipproc}   

\bibliography{juettner}

\end{document}